\documentclass[11pt,a4paper]{article}
\usepackage{jcappub}

\usepackage{amsmath}
\usepackage{amsfonts}
\usepackage{amssymb}
\usepackage{graphicx}
\usepackage{hyperref}
\usepackage{booktabs}
\usepackage{cleveref}
\usepackage{color}
\usepackage{changes}
\usepackage{subcaption}
\usepackage{enumerate}

\begin{document}

\title{Forecasts on the speed of gravitational waves at high $z$}

\author{Alexander Bonilla${}^{a}$,}
\emailAdd{abonilla@fisica.ufjf.br}

\author{Rocco D'Agostino${}^{b,c}$,}
\emailAdd{rdagostino@na.infn.it}

\author{Rafael C. Nunes${}^{d}$,}
\emailAdd{rafadcnunes@gmail.com}

\author{Jos\'e C. N. de Araujo${}^{d}$}
\emailAdd{jcarlos.dearaujo@inpe.br}

\affiliation{${}^{a}$Departamento de F\'isica, Universidade Federal de Juiz de Fora, 36036-330, Juiz de Fora, MG, Brazil.}
\affiliation{${}^{b}$Dipartimento di Fisica, Universit\`a di Napoli ``Federico II'', Via Cinthia 21, I-80126, Napoli, Italy.}
\affiliation{${}^{c}$Istituto Nazionale di Fisica Nucleare (INFN), Sezione di Napoli, Via Cinthia 9, I-80126 Napoli, Italy.}
\affiliation{${}^{d}$Divis\~ao de Astrof\'isica, Instituto Nacional de Pesquisas Espaciais, Avenida dos Astronautas 1758, S\~ao Jos\'e dos Campos, 12227-010, SP, Brazil.}

\abstract{
The observation of GW170817 binary neutron star (BNS) merger event has imposed strong bounds on the speed of gravitational waves (GWs) locally, inferring that the speed of GWs propagation is equal to the speed of light. Current GW detectors in operation will not be able to observe BNS merger to long cosmological distance, where possible cosmological corrections on the cosmic expansion history are expected to play an important role, specially for investigating possible deviations from general relativity. Future GW detectors designer projects will be able to detect many coalescences of BNS at high $z$, such as the third generation of the ground GW detector called Einstein Telescope (ET) and the space-based detector deci-hertz interferometer gravitational wave observatory (DECIGO). In this paper, we relax the condition $c_T/c = 1$ to investigate modified GW propagation where the speed of GWs propagation is not necessarily equal to the speed of light. Also, we consider the possibility for the running of the Planck mass corrections on  modified GW propagation. We parametrize both corrections in terms of an effective GW luminosity distance and we perform a forecast analysis using standard siren events from BNS mergers, within the sensitivity predicted for the ET and DECIGO. We find at high $z$ very strong forecast bounds on the running of the Planck mass, namely $\mathcal{O}(10^{-1})$ and $\mathcal{O}(10^{-2})$ from ET and DECIGO, respectively. Possible anomalies on GW propagation are bound to $|c_T/c - 1| \leq 10^{-2} \,\,\, (10^{-2})$ from ET (DECIGO), respectively. We finally discuss the consequences of our results on modified gravity phenomenology.}


\maketitle
\section{Introduction}

Astronomical information from gravitational wave (GW) observations will open a new and wide spectrum of possibilities to investigate fundamental physics, which might shed light to clarify open questions in modern cosmology, especially regarding the dark sector of the Universe. One of the most fascinating and important sources of GWs is certainly  the gravitational radiation emitted from binary neutron star (BNS) mergers. At present, one BNS merger event has been detected, the GW170817 event \cite{GW170817},  accompanied by its electromagnetic counterpart, the GRB 170817A event \cite{GRB170817A}, located at 40 Mpc ($z \approx 0.01$). This event was also the first standard siren (SS) observation, the GWs analog of astronomical standard candles, and opened the window for the multi-messenger GW astronomy (we refer the reader to \cite{LIGO_1} for a summary of all GW detections up to the present time). Although the GW170817 event is located at very low $z$, preliminary cosmological information and consequences of this observation are important to the understanding of our Universe locally. These observations were also used to measure $H_0$ at 12\% accuracy \cite{SS01}. An improvement of this result was presented in \cite{SS02, SS03}, while we refer to \cite{H0_GW1,H0_GW2,H0_GW3} for proposals to use SS to measure $H_0$ with more accuracy in the near future. A very important consequence of this BNS signals was the strong bound placed on the GW speed, $|c_T/c - 1| \leq 10^{-16}$, where $c_T$ and $c$ are the propagation speed of the GWs and light, respectively. In practical physical terms, this means that, locally, the speed of GWs propagation is equal to the speed of light. Strong constraints have been also imposed on modified gravity/dark energy scenarios \cite{GW_MG01,GW_MG02,GW_MG03,GW_MG04}. See also \cite{GW_MG05,GW_MG06,GW_MG07,GW_MG08,GW_MG09,GW_MG10} for discussion of the speed of GWs propagation on modified gravity models.

The detectability rate of the BNS merger with their electromagnetic counterpart from the current LIGO/VIRGO sensitivity is expected to be low for the next years. Also, the sensitivity of these detectors will not able to detect BNS merger at high $z$. The central importance of GW astronomy is testified by the plans for construction of several GW observatory interferometers, beyond the present performance of the LIGO and Virgo interferometers, such as Cosmic Explore \cite{CE}, Einstein Telescope (ET) \cite{ET_design01,ET_design02},
LISA \cite{LISA}, DECIGO \cite{DECIGO}, TianQin \cite{TianQin}, IMAGEM \cite{IMAGEM}, among others, to observe GWs in the most diverse frequencies bands and different types of GW sources. Certainly, the most promising projects to detect SS events from BNS merger will be the ET and DECIGO detectors, given their frequency bands and sensitivity designs. These detectors will be able to detect thousands of GWs events with great accuracy, and have been widely used for cosmological constraint investigations \cite{ET01,ET02,ET03,ET04,ET05,ET06,ET07,ET08,ET09,ET10,ET11,ET12,ET13,DECIGO01,DECIGO02,DECIGO03,DECIGO04,DECIGO05,DECIGO06,DECIGO07,DECIGO08}.

In this work, setting $c_T/c = 1$ for $z < 0.1$ (cosmological distance ten times greater than GW170817), we relax the condition $c_T/c = 1$ at high $z$ and we perform a forecast analysis on the ratio $c_T/c$, up to $z = 2$, from some modified gravity parametric models where the speed of GWs propagation is not necessarily equal to the speed of light.  Also, we consider the possibility for the running of the Planck mass corrections on modified GW propagation. To this purpose, we generated standard siren mock catalogs from BNS mergers, within the sensitivity predicted for the ET and DECIGO. 

The paper is structured as follows. In Section \ref{sec-model}, we present the theoretical scenario in which the GW propagation is modified when $c_T/c \neq 1$, as well as also in presence of the running of the Planck mass corrections. In Section \ref{Methodology}, we describe briefly our methodology and the main forecast analysis results. In Section \ref{Consequences}, we discuss consequences of our results on modified gravity models. Lastly, in Section \ref{Conclusions} we outline our final considerations and perspectives. Throughout the text, the prime symbols indicate derivatives with respect to the conformal time, and a subscript zero refers to a quantity evaluated at the present time.

\section{Gravitational Waves in Modified Gravity}
\label{sec-model}

The most general tensor metric perturbation evolution, under the FLRW metric, can be written as \cite{Saltas04}

\begin{equation}
\label{gw_ft}
h''_A + (2 + \nu) \mathcal{H} h'_A + (c_T^2 k^2 + \mu^2)h_A = \Pi_A\ ,
\end{equation}
where $h_{A}$ is the metric tensor perturbation, being $A = \{+, \times\}$ the label of the
two polarization states, and $\mathcal{H}$ is the Hubble rate in conformal time.
The quantities $\nu$, $c_T$ and $\mu$ represent the running of the effective Planck mass, the GW propagation speed and the effective graviton mass, respectively. The function $\Pi_A$ denotes extra sources generating GWs, which we assume to be null. The running of the Planck mass enters as a friction term and it is responsible for modifying the amplitude of the tensor modes, acting as a damping term during the cosmic time. This is also related to the strength of gravity. The term $c_T^2 k^2 + \mu^2$ accounts for modifications of the GW phase. In general, all these functions depend on the parameters of a specific theory (e.g., see \cite{Atsushi01, Ezquiaga_Zumalacarregui} and references therein for more details on the physical interpretation of these functions). 

Recently, it has been discussed that modifications in the underlying gravity theory can affect not only the generation mechanism (waveform of the GWs), but also their propagation through cosmological distances \cite{MG_amplitude01, MG_amplitude02}. Since the GW amplitude is inversely proportional to the luminosity distance, the modification in the amplitude and phase coming from Eq.~(\ref{gw_ft}) can be interpreted as a correction to the GW luminosity distance on general theories of modified gravity (see also \cite{Atsushi01, Atsushi02}). Assuming theories with $\mu = 0$ and for tensor modes inside the horizon, where the GW wavelength is much smaller than the cosmological horizon, we have that the effective GW luminosity distance for non-trivial function $\nu$ and $c_T$ satisfies the equation\footnote{For a didactic deduction of Eq.~(\ref{dL_Gw}), see Appendix A in \cite{LISA_MG}.} 
\begin{equation}
\label{dL_Gw}
d_L^{GW}(z) = \sqrt{\frac{c_T(z)}{c_T(0)}} \exp \left[\dfrac{1}{2} \int_0^{z} \dfrac{dz'}{1 + z'} \nu (z') \right] \times (1+z) \int_0^z \frac{c_T(z') dz'}{H(z')}\ ,
\end{equation}
where, for $\nu = 0$ and $c_T = 1$, we recover the general relativity case ($\Lambda$CDM cosmology), that is, $d_L^{GW}(z) = d_L^{EM}(z)$, where $d_L^{EM}$ is the standard luminosity distance for an electromagnetic signal.
Generalizations and interpretations of some effective GW luminosity distance have been recently studied in the context of modified gravity theories (see, e.g., \cite{MG_dL01,MG_dL02,MG_dL02_1,MG_dL03,MG_dL04,MG_dL05,MG_dL07,MG_dL08,MG_dL09,MG_dL10,MG_dL11,MG_dL12,MG_dL13}).

\begin{figure}
\begin{center}
\includegraphics[width=4in]{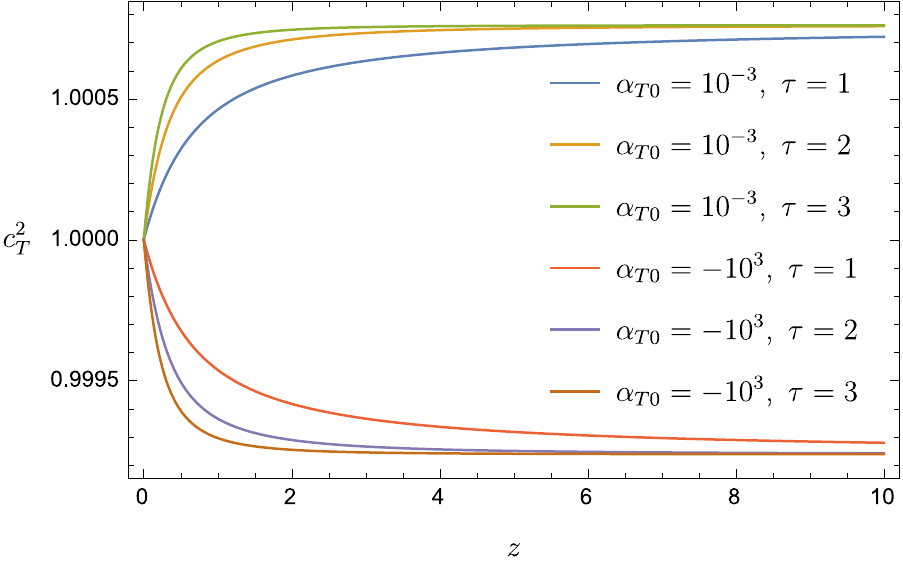} 
\caption{GW propagation speed as a function of the redshift for  the tensor speed excess parametrization defined in \ref{new_cT}, for different values of the free parameters of the theory.}
\label{cT2_new}
\end{center}
\end{figure}

In order to move on, we need to specify a gravitational model. A common procedure is to choose phenomenologically functional forms to model the dynamics of $\nu$ and $c_T$. In the recent literature, this has been done through $\alpha_i$'s functions (see, e.g., \cite{Bellini, alphai_01, alphai_02}), especially to investigate the Horndeski gravity and beyond. The quantities $\nu$ and $c_T$ are well modeled for $\alpha_M$ (Planck-mass run rate) and $\alpha_T$ (tensor speed excess), respectively, following the relationships  $\nu = \alpha_M$ and $c^2_T(z) = 1 + \alpha_T(z)$. Typically, the $\alpha_i$'s evolution is given as some power $n$ of the scale factor $a(t)$ or the dark energy density $\Omega_{de}(a)$. In the present work, we adopt two parametrizations for $\alpha_i$: 

\begin{enumerate}[I.]
\item 
First, let us consider the well-known form $\alpha_i=\alpha_{i0} a^n$, where the label $i$ refers to $M$ and $T$. Denoting by $n_1$ and $n_2$ the powers over the Planck-mass run rate and tensor speed excess, respectively, we thus have $\alpha_M = \alpha_{M0} a^{n_1}$ and $\alpha_T = \alpha_{T0} a^{n_2}$. We will refer to this scenario as \textit{model I}.

\item 
In order to check for possible dependence on the parametrization, we propose a new $\alpha_T$ parametrization, which is characterized by a smoother parametric dynamics with respect to the previous model:

\begin{equation}
\label{new_cT}
\alpha_T = \alpha_{T0} \tanh(1-a^\tau)\ ,
\end{equation}
where $\tau > 0$ is a constant. Our parametrization, which will be called \textit{model II} in what follows, presents some advantages in comparison with model I. The $\tau$ parameter controls the shape of the cosmic evolution of $c_T(z)$, determining the possible amount of variation from the speed of light at high $z$. For positive  values of $\tau$, $c_T$ asymptotically tends to $c$ at low $z$, and the higher the value of $\tau$, the faster $c_T \rightarrow c$ at small cosmological distances, while $c_T/c \neq 1$ at high $z$ in such a way to be constant for $z\gg 1$. In Figure \ref{cT2_new}, we quantify these effects for reasonable values of $\alpha_{T0}$ and $\tau$. In this scenario, we keep $\alpha_M = \alpha_{M0} a^{n_1}$ for the running of the Planck mass.
\end{enumerate}

\noindent An important role is played by the stability conditions of the theory. Suitable values of the free parameters must be considered to have a stable theory throughout the evolution of the Universe (see \cite{alphai_02} and reference therein). The stability conditions for $\alpha_M =\alpha_{M0} a^{n_1}$ can be summarized as follows:
\begin{align}
&n_1 > \dfrac{5}{2}, \ \alpha_{M0} < 0\ ; \label{stab1}\\
&0 < n_1 < 1 + \dfrac{3\Omega_{m0}}{2}, \ \alpha_{M0} > 0 \ , \label{stab2}
\end{align}
where $\Omega_{m0}$ is the present normalized matter density. Moreover, the stability of the tensor modes requires $c^2_T(z) = 1 + \alpha_T(z) > 0$. In the following, we assume these conditions in our results for both models.

\begin{figure}
\begin{center}
\includegraphics[width=3.in]{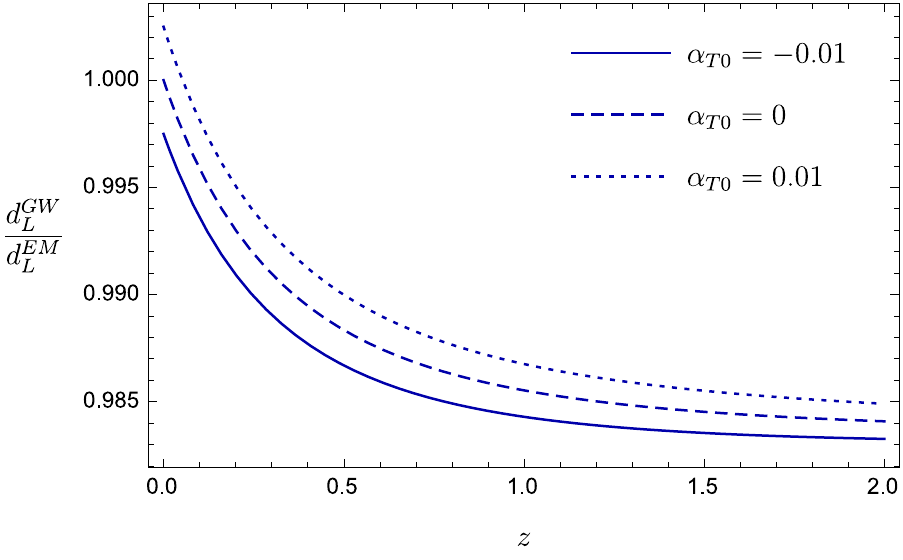} 
\includegraphics[width=3.in]{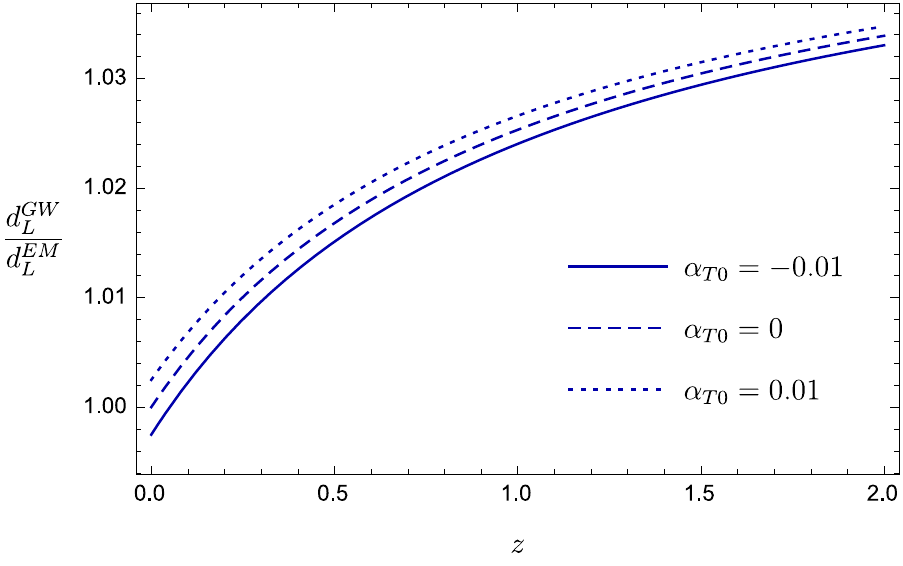}
\caption{Corrections on the effective GW luminosity distance (cf. Eq.~(\ref{dL_Gw})) as a function of the redshift for \textit{model I} with different values of the tensor speed excess $\alpha_{T0}$ with fixed values of $\alpha_{M0}$. Left panel: $\alpha_{M0}=-0.1$, $n_1 = 3$, $n_2=1$. Right panel: $\alpha_{M0}=0.1$, $n_1 = 1$, $n_2=1$. The limit $d_L^{EM}(z)/d_L^{EM} = 1$ represents general relativity.}
\label{fig:dL}
\end{center}
\end{figure}

\begin{figure}
\begin{center}
\includegraphics[width=3.in]{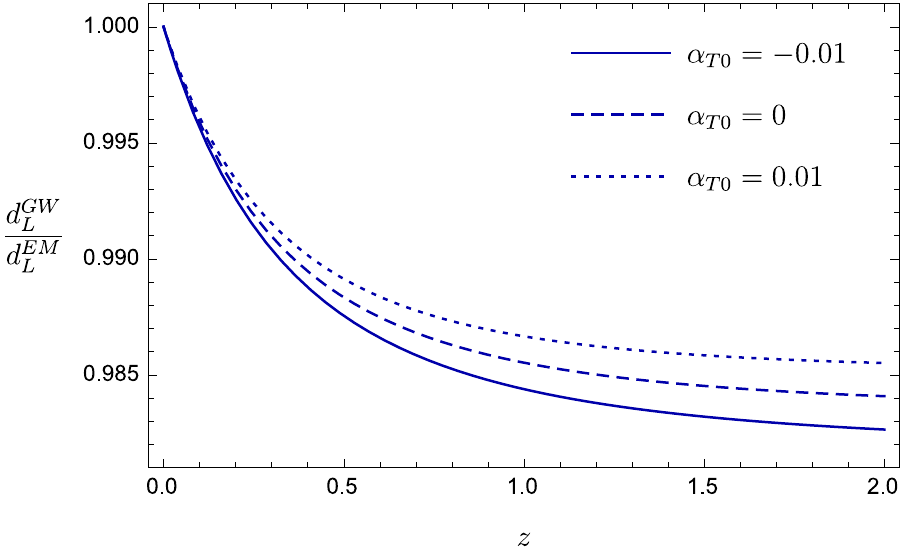} 
\includegraphics[width=3.in]{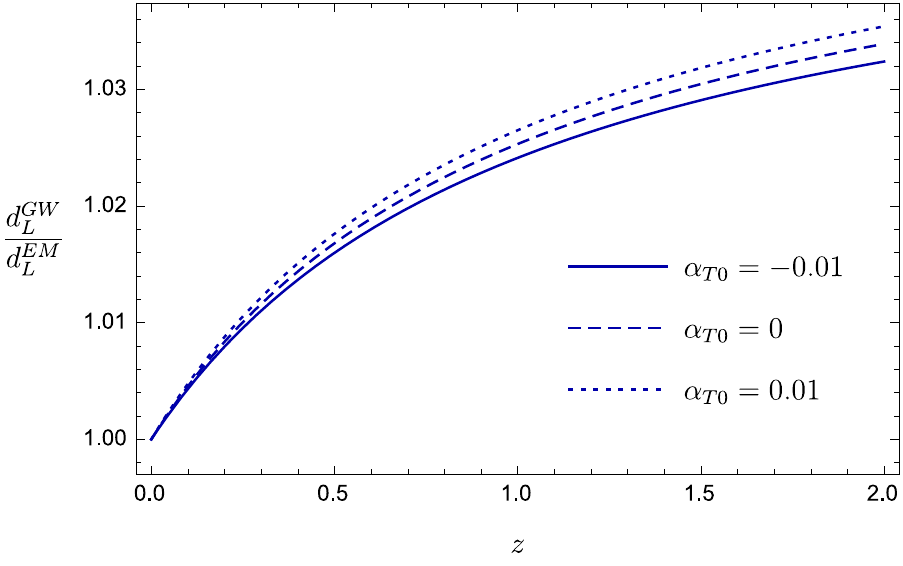}
\caption{Corrections on the effective GW luminosity distance taking into account the new parametrization (\ref{new_cT}), \emph{i.e.} \textit{model II}, for different values of the tensor speed excess $\alpha_{T0}$ with fixed values of $\alpha_{M0}$. Left panel: $\alpha_{M0}=-0.1$, $n_1 = 3$, $\tau=1$. Right panel: $\alpha_{M0}=0.1$, $n_1 = 1$, $\tau=1$.}
\label{fig:dL2}
\end{center}
\end{figure}

In Figure \ref{fig:dL}, we show the corrections due to model I on the effective GW luminosity distance induced from both contributions $\alpha_{T0}$ and $\alpha_{M0}$, inducing $d_L^{GW} > d_L^{EM}$ and $d_L^{GW} < d_L^{EM}$ for $\alpha_{M0} > 0$ and $\alpha_{M0}< 0$, respectively. In drawing the plot, we keep $\alpha_{M0}$ fixed and we vary $\alpha_{T0}$ between [-0.01, 0.01] for both stability conditions $\alpha_{M0} > 0, \,\, < 0$. We note that for $\alpha_{M0} < 0$, corrections with positive (negative) values of $\alpha_{T0}$ will induce $d_L^{GW} > d_L^{EM}$ ($d_L^{GW} < d_L^{EM}$), compared to the $\alpha_{T0} = 0$ prevision. While, when considering $\alpha_{M0} > 0$, for $\alpha_{T0} > 0 \, (< 0)$  we have $d_L^{GW} > d_L^{EM}$ ($d_L^{GW} < d_L^{EM}$). Therefore, the determination of the signal, imposed by the stability conditions, provides a variety of possible corrections on the effective luminosity distance. 

In Figure \ref{fig:dL2}, we show the corrections on $d_L^{GW}/d_L^{EM}$ predicted by model II. We soon notice different properties in our proposal. In all the cases, considering reasonable values of all free parameters of the theory ($\alpha_T$ and $\alpha_M$ functions), we find the convergence $d_L^{GW}/d_L^{EM} \rightarrow 1$ at low $z$ and corrections $d_L^{GW} \neq d_L^{EM}$ at high $z$. The presence of a possible time variation of the Planck-mass will determine $d_L^{GW} < d_L^{EM}$ (when $\alpha_{M0} < 0$) or $d_L^{GW} > d_L^{EM}$ (when $\alpha_{M0} > 0$). The sign determination of $\alpha_{T0}$ will mitigate possible corrections $d_L^{GW} \neq d_L^{EM}$ at high $z$ (see Figure \ref{fig:dL2}).

In the following, we will forecast bounds on the free parameters for all the scenarios using mock catalogs of the standard siren events from BNS mergers.

\section{Methodology and Results}
\label{Methodology}

We shall now analyze the waveform emitted by the binary system. For a given a GW strain signal $h(t) = A(t) \cos [\Phi(t)]$, one can use the stationary-phase approximation for the orbital phase of inspiraling binary system to obtain its Fourier transform $\tilde{h}(f)$. In the case of a coalescing binary system of masses $m_1$ and $m_2$, we have
\begin{equation}
\label{waveform}
\tilde{h}(f) = Q \mathcal{A} f^{-7/6} e^{i\Phi(f)}\ ,
\end{equation}
where $\mathcal{A} \propto 1/d_L^{GW}$ is the modified amplitude given by Eq.~(\ref{dL_Gw}). The $\Phi(f)$ is the inspiral phase of the binary system. More details on the post-Newtonian coefficients and waveforms can be found in \cite{MG_dL04} and references therein.

Once the modified GW signal propagation is defined, for a high enough signal-to-noise ratio (SNR), we can obtain upper bounds on the free parameters of the GW signal $\tilde{h}(f)$ by means of the Fisher information analysis. Estimating $d_L(z)$ from GW standard sirens mock data is a well consolidated methodology, and we refer to \cite{MG_dL04} for a detailed description. Pioneer studies in this regard are represented by the works \cite{Schutz, Holz}. In what follows, we briefly describe our methodology.

\begin{figure}
\begin{center}
\includegraphics[width=4.8in]{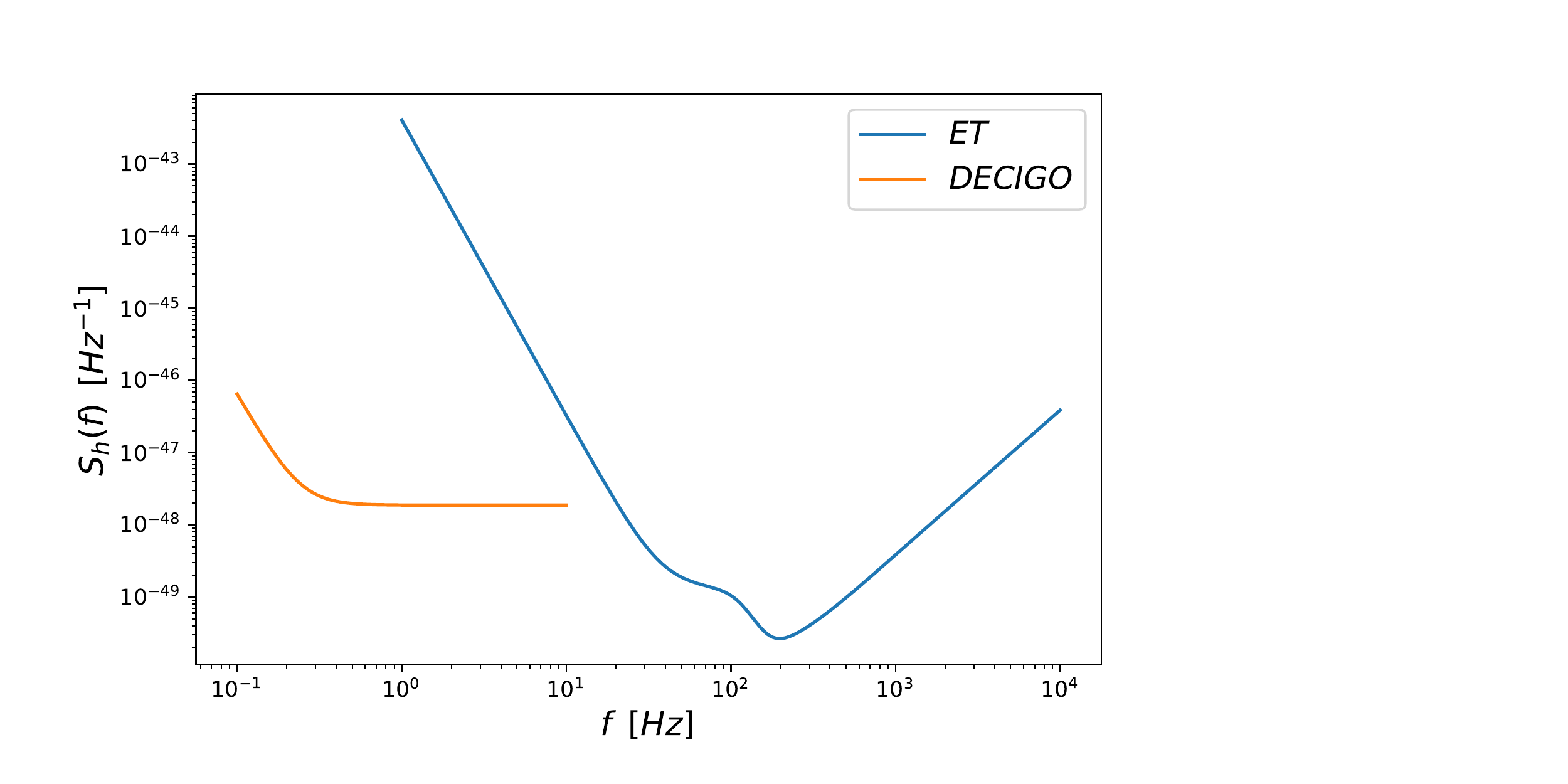} 
\caption{Detector spectral noise density for the
Einstein Telescope (ET) and deci-hertz interferometer gravitational wave observatory (DECIGO).}
\label{fig:PSD}
\end{center}
\end{figure}

We considered the ET and DECIGO power spectral density noises to generate our mock standard siren catalogs. Figure \ref{fig:PSD} shows the spectral noise density curve of both experiments. ET is a third-generation ground detector, covering the frequency range $1-10^4$ Hz. The signal amplitude which ET is sensitive to is expected to be ten times larger than the current advanced ground-based detectors.  
The ET conceptual design study predicts an order of $10^3-10^7$ BNS detections per year. Nevertheless, only a small fraction ($\sim 10^{-3}$) of them is expected to be accompanied by a short $\gamma$-ray burst observation. Assuming a detection rate of $\mathcal{O}(10^5)$, the events with short $\gamma$-ray bursts will be $\mathcal{O}(10^2)$ per year. DECIGO is the most sensitive GW detector proposed in the $0.1-10$ Hz band, enough to detect a cosmological GW background generated at early times, the mergers of intermediate-mass black holes and a large number of BNS merges. These GW sources enable us to measure the cosmological expansion with unprecedented precision \cite{DECIGO01, DECIGO02, DECIGO03, DECIGO04, DECIGO05, DECIGO06} and to test alternative theories of gravity \cite{Abedi18,DAgostino18,Capozziello19}. Based on this setup, in \cite{DECIGO05} it has been shown that cosmological parameters can be accurately measured by DECIGO with a precision of $\sim$ 1\% assuming a large number of BNS.

In our simulations, we considered 1000 BNS mock GW standard sirens merger events, up to $z = 2$, for the ET and DECIGO detectors. The redshift distribution of the sources, taking into account evolution and stellar synthesis, is well described by
\begin{eqnarray}
P(z) \propto \frac{4 \pi \chi^2(z) R(z)}{H(z)(1+z)}\ ,
\end{eqnarray}
where $\chi$ is the comoving distance and $R(z)$ describes the time evolution of the burst rate and is given by $R(z)=1 + 2z$ for $z<1$, $R(z) = 3/4(5-z)$ for $1 \leq z \leq 5$, and $R(z)=0$ for $z>5$.
The input masses of the NSs in our mock catalogs are randomly sampled from uniform distributions within $[1-2]~{\rm M}_{\odot}$. When generating our mock GW events, we have only considered mergers with ${\rm SNR}>8$\footnote{A relevant quantity for generating the mock catalogue is the signal-to-noise ratio (SNR) associated with each simulated event. The SNR is given by:
\begin{eqnarray}
{\rm SNR}^2 \equiv 4 Re \int_{f_{\min}}^{f_{\max}} df\, \frac{\vert h(f)\vert ^2}{S_n}\,,
\label{eq:snr}
\end{eqnarray}
where $S_n(f)$ is the spectral noise density. The upper cutoff frequency $f_{\max}$ is determined by the last stable orbit. The lower cutoff frequency $f_{\min}$ is instead determined for each specific GWs detector.}.

The redshift range and event numbers rate are fully compatible with the sensitivity of both instruments. For each event, in each catalogue, we estimate the measurement error on the luminosity distance for the ET and DECIGO configurations by applying the Fisher matrix analysis on the waveforms (see \cite{MG_dL04} for details). We then construct the Fisher matrix for the parameters of the cosmological model under consideration as

\begin{equation}
\label{Fisher}
\begin{aligned}
F_{ij} = \sum_n \frac{1}{\sigma^2_{\rm ins} + \sigma^2_{\rm lens}(z_n) + \sigma^2_{v}(z_n)} \frac{\partial d_L(z_n)}{\partial \theta_i} \frac{\partial d_L(z_n)}{\partial \theta_j}\ ,
\end{aligned}
\end{equation}
where the sum $n$ runs over all standard sirens mock events. The derivatives are performed with respect to the cosmological parameters $\theta_i = \{H_0, \,\, \Omega_{m0}, \,\, \alpha_{M0}, \,\, \alpha_{T0}, \,\, n_1, \,\, n_2 \, (\tau) \}$ evaluated at their fiducial input values. In our analysis, we used $\theta_i= \{67.4, \,\, 0.30, \,\, 0.0, \,\, 0.0, \,\, 3.0 \,\, (1.0), \,\, 1.0 \,\, (1.0) \}$ as fiducial values. The $n_1$ values have been chosen to fulfil the stability conditions (\ref{stab1})-(\ref{stab2}) from $\alpha_{M0} < 0 \,\, (> 0)$. The quantities $\sigma^2_{\text{ins}}$, $\sigma^2_{\rm lens}(z_n)$ and $\sigma^2_{v}(z_n)$ are the instrumental, lensing and galaxy peculiar velocity errors, respectively. For both instruments, we consider the galaxy peculiar velocity error given by \cite{Gordon}

\begin{equation}
\label{error_galaxy}
\begin{aligned}
\sigma^2_{v} = d_L \times \Big|1 - \frac{(1+z)^2}{H(z) d_L} \Big| \sigma_{\rm v, gal}\ ,
\end{aligned}
\end{equation}
where $\sigma_{\rm v, gal}$ is the one-dimensional velocity dispersion of the galaxy, set to be 300 km/s, as a rough estimate. It is well known that
systematic effects due $\sigma^2_{v}$ are relevant only at very low $z$. In Appendix \ref{appendix}, we show how much different $\sigma_{\rm v, gal}$ values can affect our simulations. As we are interested in simulated data at high $z$, the effects produced by $\sigma_{\rm v, gal}$ are minimal in our forecast analysis. Thus, without loss of generality, let us assume 300 km/s in all the forecast analyses performed in this work.

For ET, the other uncertainties can be summarized as \cite{ET02}
\begin{equation}
\label{error_ET}
\begin{aligned}
\sigma^2_{\rm ins} + \sigma^2_{\rm lens} = \Big(\frac{2 d_L}{\rm SNR} \Big)^2 + (0.05 z d_L)^2\   ,
\end{aligned}
\end{equation}
while, for DECIGO, we have \cite{Barausse}
\begin{equation}
\label{lensing_DECIGO}
\sigma^2_{\rm ins} + \sigma^2_{\rm lens} = \Big(\frac{2 d_L}{\rm SNR} \Big)^2 +d_L \times 0.066 \Big[ \frac{1 - (1+z)^{-0.25}}{0.25}  \Big]^{1.8}.
\end{equation}
Thus, the total uncertainty on the measurement for each experiment is given by 
\begin{equation}
\sigma_{d_L} = \sqrt{\sigma^2_{\rm ins} + \sigma^2_{\rm lens} + \sigma^2_{v}}\ .
\end{equation}

We calculated the SNR of each event and confirmed that it is a GW detection if SNR $>$ 8. When performing the integration, we assumed $f_{low} = 1$ ($0.1$) Hz for ET (DECIGO). As for the upper frequency limit, we considered  $f_{upper} = 2 f_{LSO}$ for both cases, where $f_{LSO} = 1/(6^{3/2} 2 \pi M_z)$ is the orbital frequency at the last stable orbit, with $M_z = (1 + z)M$. In the DECIGO case, if $f_{upper} > 10$ Hz for any event, then we fixed $f_{upper} = 10$ Hz. More details about NSs parameters distributions can be found in \cite{MG_dL04}.
\\

\begin{table}[h!]
\small
\begin{center}
\begin{tabular}{c| c  c| c c }
\hline
\hline
& & Stability condition $\alpha_{M0}<0$ & Stability condition $\alpha_{M0} > 0$ \\
\hline
Parameter & $\sigma$(ET) & $\sigma$(DECIGO) & $\sigma$(ET) & $\sigma$(DECIGO) \\
\hline
$\alpha_{T0}$  & 0.053 &  0.015 &  0.057 &  0.016  \\
$\alpha_{M0}$  & $>$ -0.18 &   $>$ -0.050  & $<$ 0.087 & $<$ 0.052   \\
$n_{1}$        & 0.41 &  0.13  & 0.33 &  0.12 \\
$n_{2}$        & 0.18 &  0.053  & 0.17 &  0.052 \\
\hline
\hline
\end{tabular}
\caption{Forecast constraints from the ET and DECIGO experiments for \textit{model I} under both stability conditions $\alpha_{M0}<0, >0$. The notations $\sigma$(ET) and $\sigma$(DECIGO) represent
the 95\% C.L. uncertainties on the fiducial input values.}
\label{tab:results_Model_I}
\end{center}
\end{table}

\begin{table}[h!]
\small
\begin{center}
\begin{tabular}{c| c  c| c c }
\hline
\hline
& & Stability condition $\alpha_{M0}<0$ & Stability condition $\alpha_{M0} > 0$ \\
\hline
Parameter & $\sigma$(ET) & $\sigma$(DECIGO) & $\sigma$(ET) & $\sigma$(DECIGO) \\
\hline
$\alpha_{T0}$  & 0.11 &  0.036 &  0.11 &  0.036  \\
$\alpha_{M0}$  & $>$ -0.16 &   $>$ -0.050  & $<$ 0.079 & $<$ 0.023   \\
$n_{1} $       & 0.41 &  0.13  & 0.34 &  0.11 \\
$\tau$        & 0.21 &  0.047  & 0.21 &  0.048 \\
\hline
\hline
\end{tabular}
\caption{Forecast constraints from the ET and DECIGO experiments for \textit{model II} under both the stability conditions $\alpha_{M0}<0, >0$. The notation is the same as in Table \ref{tab:results_Model_I}.}
\label{tab:results_Model_II}
\end{center}
\end{table}

\begin{figure*}
\begin{center}
\includegraphics[width=5in]{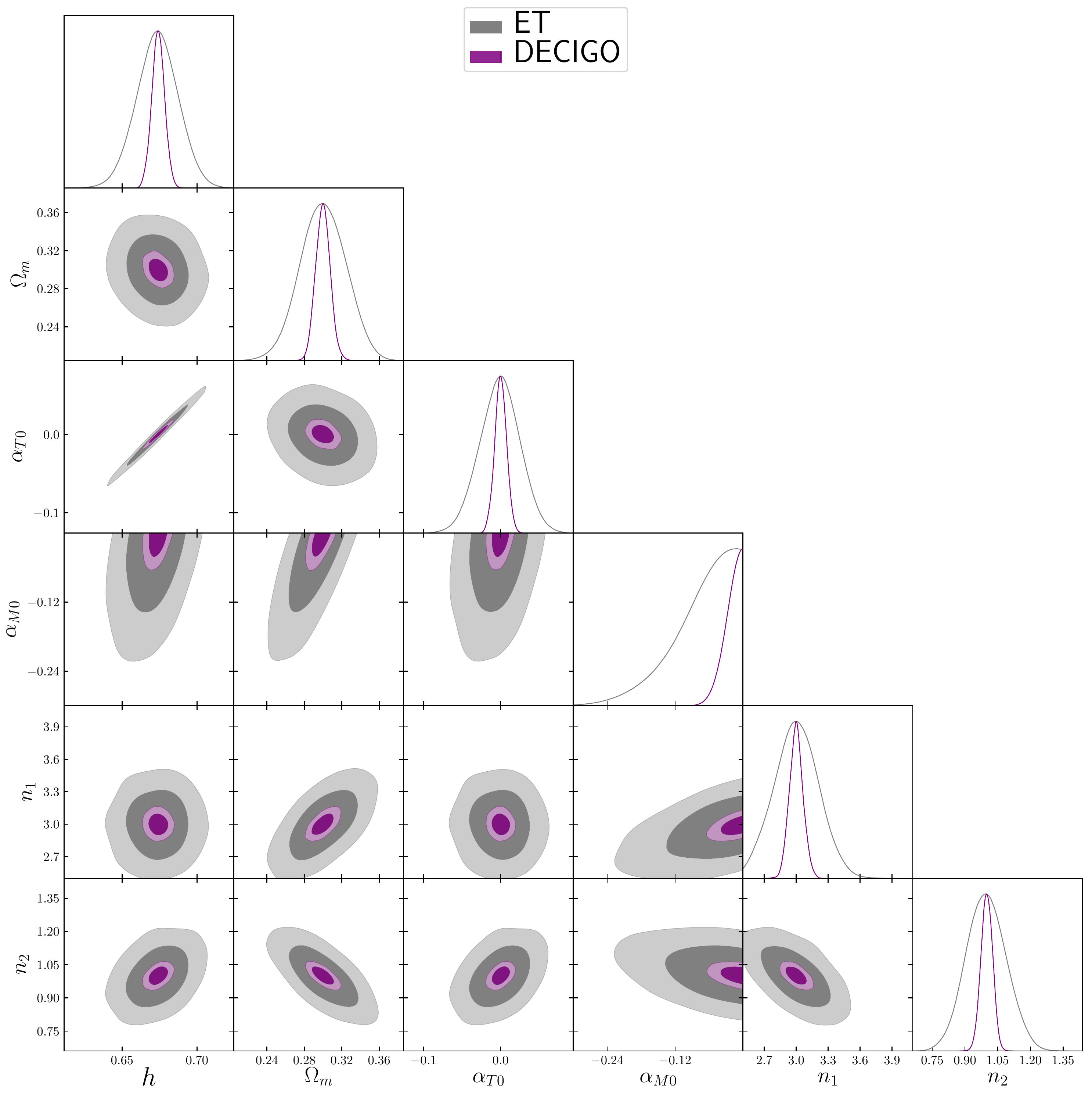}
\caption{One-dimensional marginalized distribution and 68\% and 95\% C.L. forecasted regions for the parameters of \textit{model I} from the ET and DECIGO experiments. The stability condition $\alpha_{M0}<0$ is considered.}
\label{fig:contours_alphaM0<0_Model_I}
\end{center}
\end{figure*}

\begin{figure*}
\begin{center}
\includegraphics[width=5in]{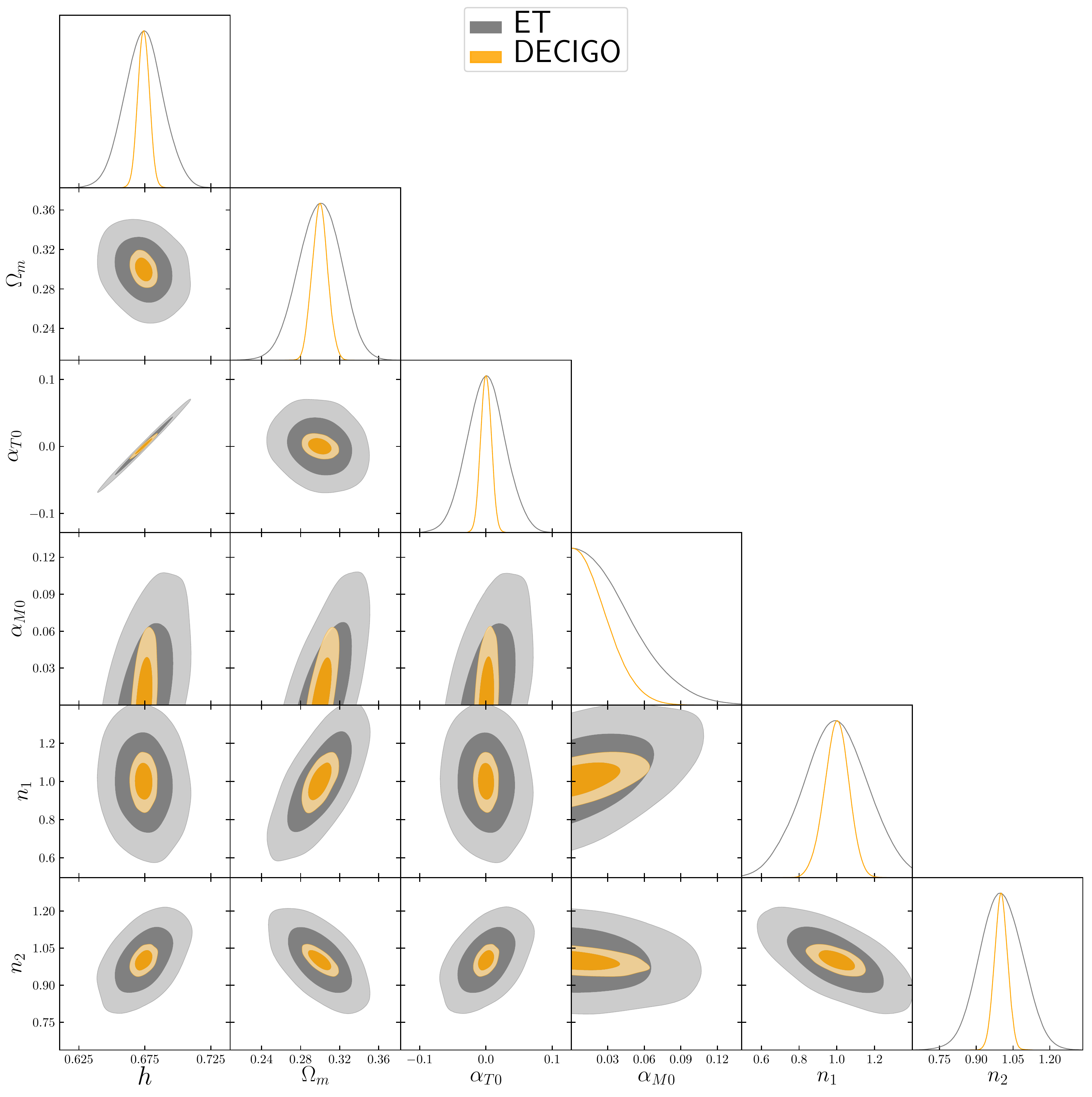}
\caption{One-dimensional marginalized distribution and 68\% and 95\% C.L. forecasted regions for the parameters of \textit{model I} from the ET and DECIGO experiments. The stability condition $\alpha_{M0}>0$ is considered.}
\label{fig:contours_alphaM0>0_Model_I}
\end{center}
\end{figure*}

\begin{figure*}
\begin{center}
\includegraphics[width=5in]{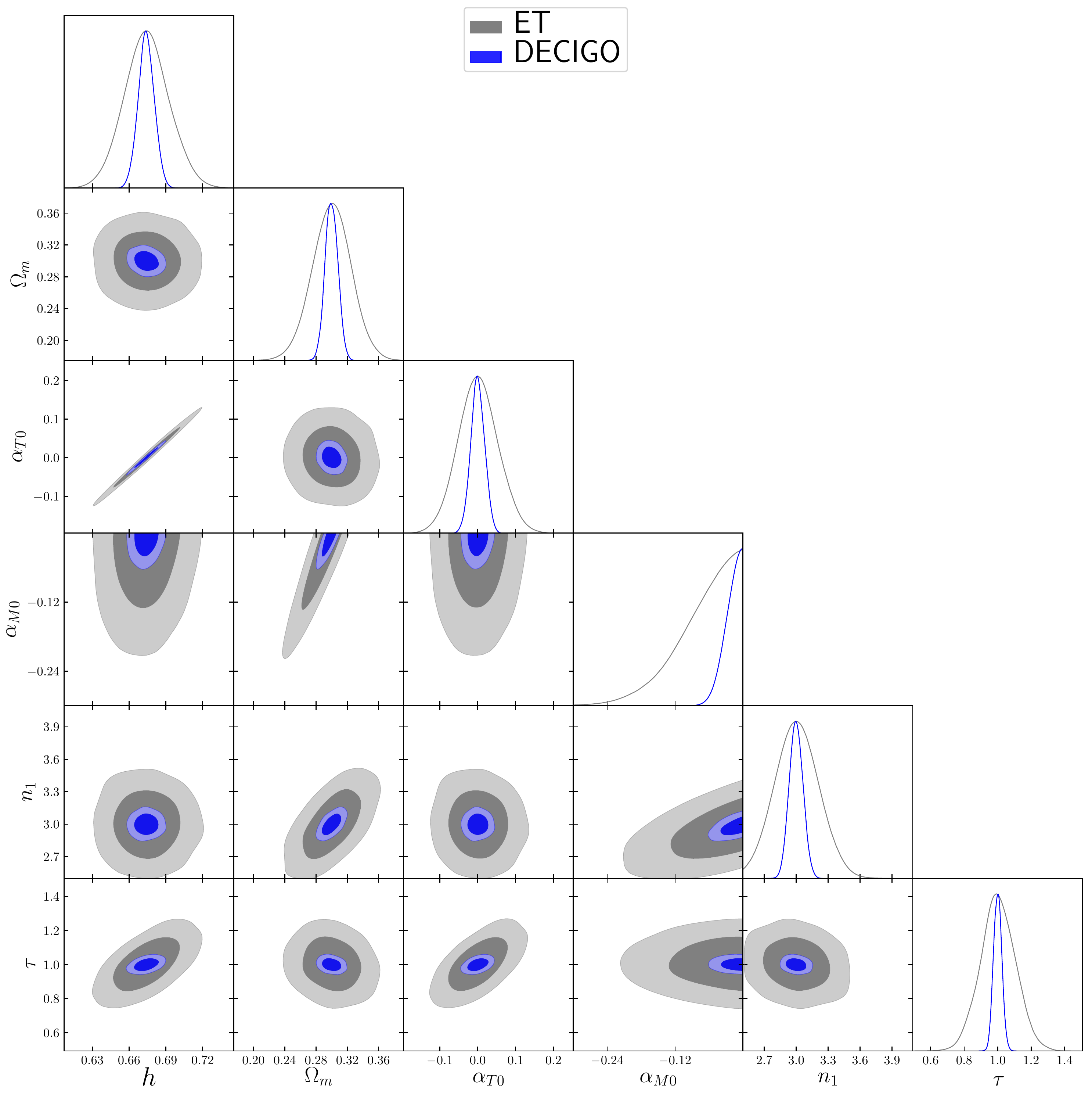}
\caption{One-dimensional marginalized distribution and 68\% and 95\% C.L. forecasted regions for the parameters of \textit{model II} from the ET and DECIGO experiments. The stability condition $\alpha_{M0}<0$ is considered.}
\label{fig:contours_alphaM0<0_Model_II}
\end{center}
\end{figure*}

\begin{figure*}
\begin{center}
\includegraphics[width=5in]{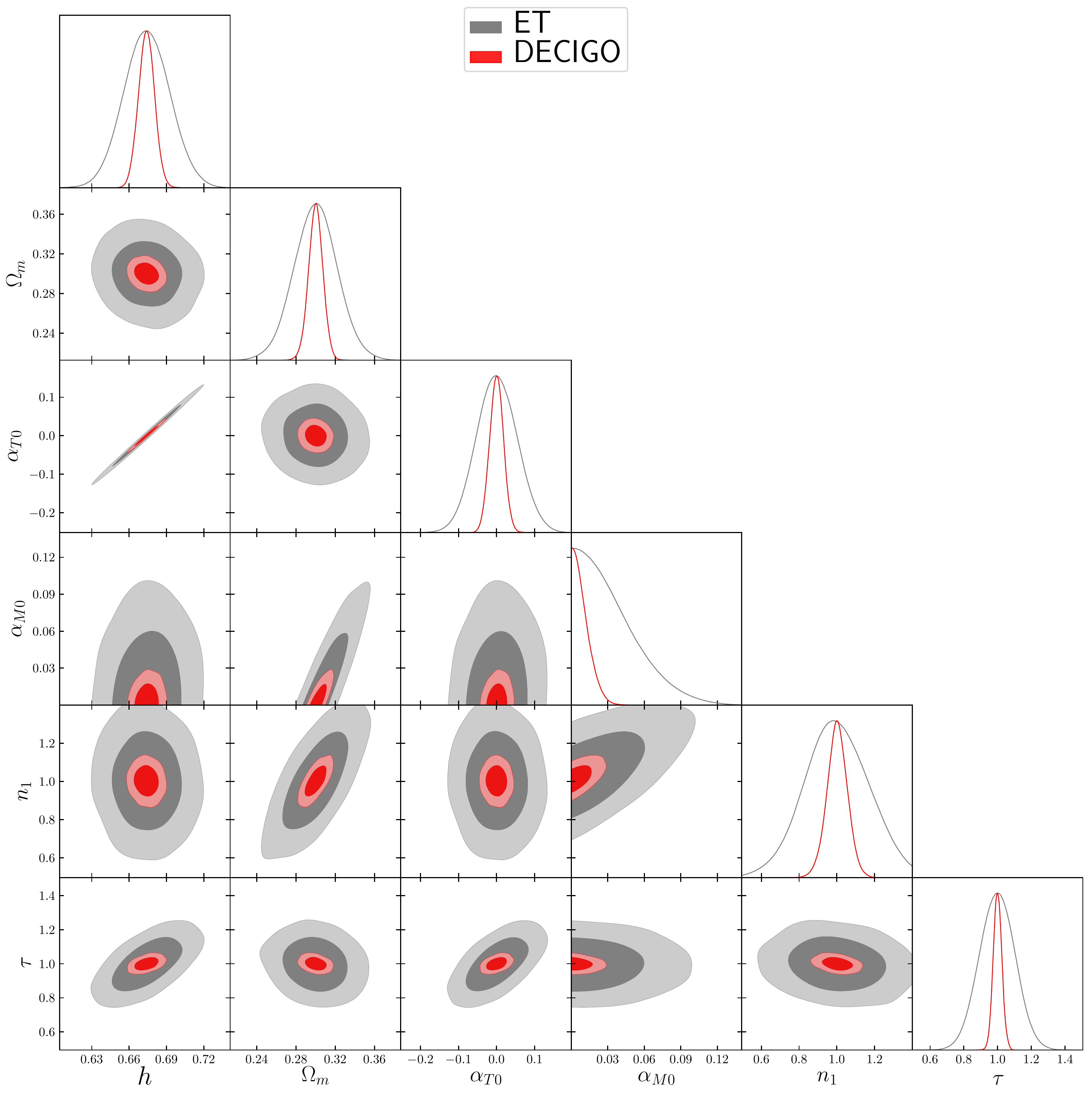}
\caption{One-dimensional marginalized distribution and 68\% and 95\% C.L. forecasted regions for the parameters of \textit{model II} from the ET and DECIGO experiments. The stability condition $\alpha_{M0}>0$ is considered.}
\label{fig:contours_alphaM0>0_Model_II}
\end{center}
\end{figure*}

Table \ref{tab:results_Model_I} and \ref{tab:results_Model_II} summarize the forecast constraints at the 95\% confidence level (C.L.) on the parameters of the theory for  models I and II, respectively. Figures \ref{fig:contours_alphaM0<0_Model_I} and \ref{fig:contours_alphaM0>0_Model_I} shows the full parameter space from both stability conditions on the running of the effective Planck mass, where a direct comparison on each parameter from each experiment can be seen for model I. Similarly, in Figures \ref{fig:contours_alphaM0<0_Model_II} and \ref{fig:contours_alphaM0>0_Model_II}, we show the parameter space concerning model II.

We are particularly interested in the limits on $\alpha_{M0}$ and $\alpha_{T0}$, especially on $\alpha_{T0}$, to quantify deviations on the speed of GW propagation. Thus, let us discuss in detail our results on these parameters.  
For all scenarios, we note that the observational bonds from DECIGO sensitivity are significantly improved when compared to ET. In fact, this is expected as DECIGO is more sensitive in amplitude than ET (cf. Figure~\ref{fig:PSD}) at low frequencies, where the binary system is still emitting signals, thus making possible to detect BNS mergers events with higher SNR values with respect to ET. Therefore, the errors measurement of the free parameters of the theory will be more accurate in the DECIGO experiment due to the higher SNR associated with the events in this band. This can be quantified analyzing the Figure of Merit (FoM) for the estimated sensitivities of the parameters $\alpha_{T0}$ and $\alpha_{M0}$ on its corresponding plane as
\begin{equation}
\label{FoM}
\begin{aligned}
\text{FoM}(M) = \vert F \vert^{1/2} \frac{\Gamma(M/2+1)}{\pi^{M/2}} (\delta \chi^2)^{-M/2}\ ,
\end{aligned}
\end{equation}
where
\begin{equation}
\label{DelChi}
\begin{aligned}
\delta \chi^2(M,n)=2\ , \quad \mathcal{G} \left[ \frac{M}{2},\ 1-\text{erf} \left( \frac{n}{\sqrt{2}} \right) \right],
\end{aligned}
\end{equation}
being $\mathcal{G}$ the inverse of the $\Gamma$ regularized function. Here, $M$ is the number of parameters of the theory, ${\rm erf}$ is the error function, $n$ is the level of statistical confidence desired, and $\vert F \vert$ is the determinant of Fisher Matrix $F_{ij}$ \cite{Press}. 
The above definition offers the advantage to make useful comparisons for different dark energy experiments or, in this case, for future GW experiments with respect to their sensitivity in constraining different cosmological parameters. Higher values of the FoM correspond to tighter constraints on model I and model II. For model I, specifically in the case of $\alpha_{M0}<0$, we find that the sensitivities on ($\alpha_{T0}, \alpha_{M0}$) from the DECIGO experiment $(\text{FoM}=40.941)$ improves the FoM from the ET experiment $(\text{FoM}=3.498)$ by a factor of $ \sim 11.704$, while, in the case with $\alpha_{M0}>0$, we find $\text{FoM}=27.221$ and $\text{FoM}=5.163$ from the DECIGO and ET experiments respectively, with an improvement by a factor of $\sim 5.272$ (see Figures \ref{fig:contours_alphaM0>0_Model_I} and \ref{fig:contours_alphaM0<0_Model_I}). For model II, we found the following results for the same set of parametric spaces: for $\alpha_{M0}<0$, DECIGO experiment $(\text{FoM}=22.052)$ improves the FoM from the ET experiment $(\text{FoM}=2.683)$ by a factor of $ \sim 8.219$ and for the case with $\alpha_{M0}>0$, we found $\text{FoM}=33.166$ and $\text{FoM}=3.932$ from the DECIGO and ET experiments respectively, with an improvement of $\sim 8.434$ (Figures \ref{fig:contours_alphaM0>0_Model_II} and \ref{fig:contours_alphaM0<0_Model_II}). This represents a great performance on the whole baseline of the scenarios under consideration.

Our forecast analysis strongly bounds possible deviations from general relativity. This is evident from the forecast bounds on the running of the Planck mass, where for the model I, we find: $-0.18 \ (- 0.05) < \alpha_{M0}\leq 0$ from ET (DECIGO) and $0\leq \alpha_{M0}<0.087\ (0.052)$ from ET (DECIGO), under the conditions $\alpha_{M0}<0$ and $\alpha_{M0}>0$, respectively. In the case of model II, we find: $-0.16 \ (- 0.05) < \alpha_{M0}\leq 0$ from ET (DECIGO) and $0\leq \alpha_{M0}<0.079\ (0.023)$ from ET (DECIGO), respectively.
These limits appear significantly more robust than the current constraints \cite{Horndeski_constraints_02,Horndeski_constraints_04,Horndeski_constraints_05}, being our results able to foresee strongly bounds on some phenomenological gravity models.

The main aim of this work is to find new limits on $\alpha_{T0}$ and, as expected, we obtained $|\alpha_{T0}|\ll 1$. In particular, we find that $\alpha_{T0}$ can be tightly constrained when analyzed by DECIGO, with an improvement of one order of magnitude with respect to ET.
It is important to emphasize that, to the authors' knowledge, the forecast analysis on $\alpha_{T0}$ has never been performed in the literature from the point of view of standard sirens GWs data. Therefore, we shall discuss two different perspectives in this context, from two different future GW experiments. 

For model I with $\alpha_{M0} < 0$, we find $\alpha_{T0} \sim  \mathcal{O}(10^{-2})$, $\mathcal{O}(10^{-2})$ from ET and DECIGO, respectively. When considering $\alpha_{M0} > 0$, we note that the forecast error on $\alpha_{T0}$ are similar to when assuming $\alpha_{M0} < 0$. This shows that the forecast sensitivity on $\alpha_{T0}$ within this parametrization is minimally dependent on the stability conditions induced from the Planck-mass run rate.

Assuming our proposal, \textit{i.e.} model II, we also find that the forecast on $\alpha_{T0}$  minimally depends on  the sign of $\alpha_{M0}$. On the other hand, we note a difference on the tensor speed excess with forecast sensitivity given by $\alpha_{T0} \sim  \mathcal{O}(10^{-1})$, $\mathcal{O}(10^{-2})$ from ET and DECIGO, respectively: the DECIGO sensitivity can improve up to one order of magnitude in comparison to ET. Apparently, different parametrizations can reasonably change the simulations (the forecast analysis) on the $\alpha_{T0}$ parameter. Due to this possible bias dependence in the theoretical input, we shall argue in the following why our phenomenological proposal is significantly more appropriate to describe the evolution of $c_T(z)$.

It is interesting to compare our simulations with results obtained by other authors. In \cite{Alonso}, combining future missions of stage IV photometric redshift and CMB surveys, the authors found $\alpha_{T0} \sim  \mathcal{O}(10^{-1})$. Thus, a direct comparison between these and our estimates reveals that forthcoming GWs data have the potential to constrain $\alpha_{T0}$ with an accuracy up to one orders of magnitude greater when compared to future CMB and LSS surveys. Proposals to measure the speed of GWs in the pre-recombination era are presented in \cite{Jeong_Kamionkowski}.

Let us now look at measurements from already catalogued data. In \cite{cT_constraints_01} combinations of various data, including Planck data, showed that from CMB is possible to obtain $\alpha_{T0} \sim  \mathcal{O}(10^{-1})$, and with significant preference for $\alpha_{T0} < 0$ from a robust joint analysis with multiple data (see table 4 there). In \cite{cT_constraints_02}, the authors find $0.55 < c_T/c < 1.42$ from  the time delay between GWs signals arriving at widely separated detectors.

Therefore, standard sirens merger events from the ET perspectives allow to get bounds on $\alpha_{T0}$ compatible with the limits obtained via current and future CMB and LSS data, and these bounds can be improved up to one order of magnitude using the DECIGO design. On the other hand, it is important to note that all of these bounds on possible anomalies in GWs propagation quantified at high $z$ are several orders of magnitude larger than those provided by the GW170817 event locally, where in fact no corrections on general relativity should be expected. Hence, investigating $c_T(z)$ at very high $z$ can open new perspectives in this sense, to test gravity models where in principle $c_T/c \neq 1$.

\begin{figure*}
\begin{center}
\includegraphics[width=3.in]{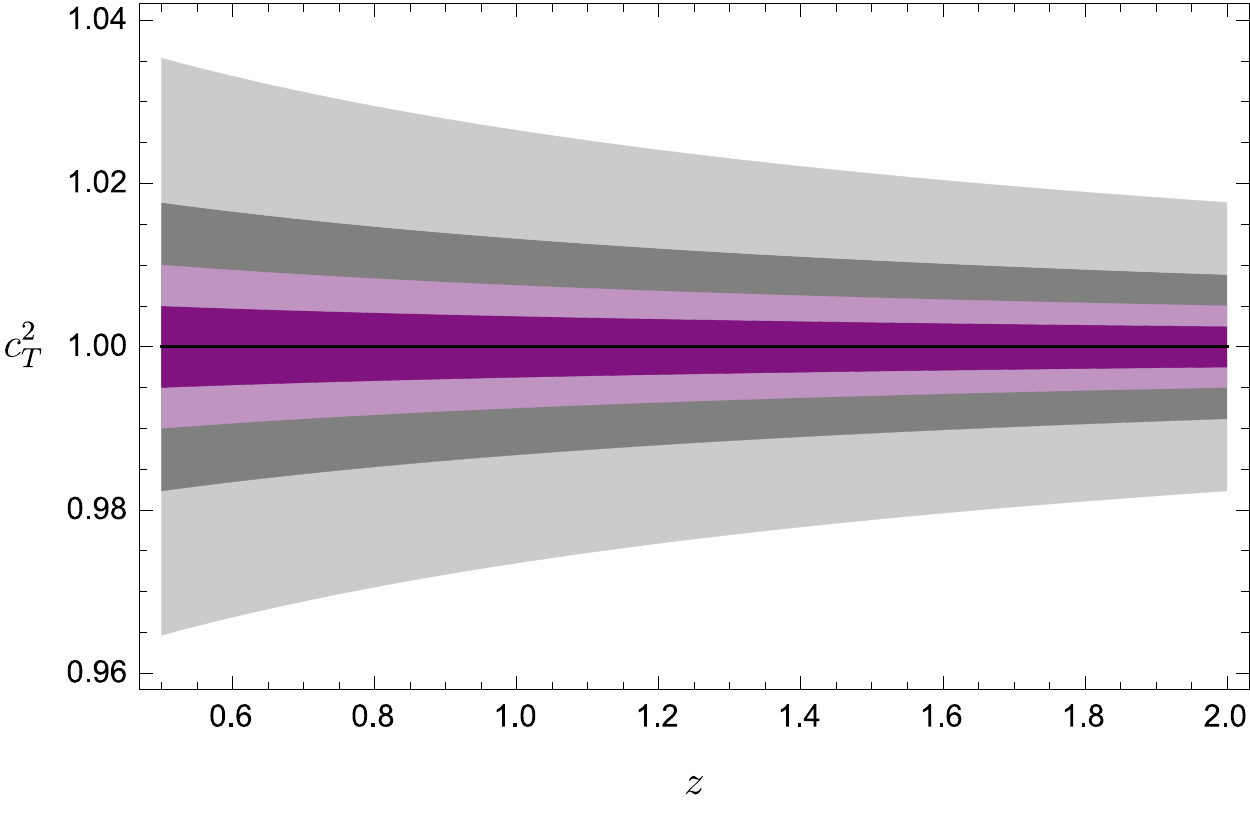} 
\includegraphics[width=3.in]{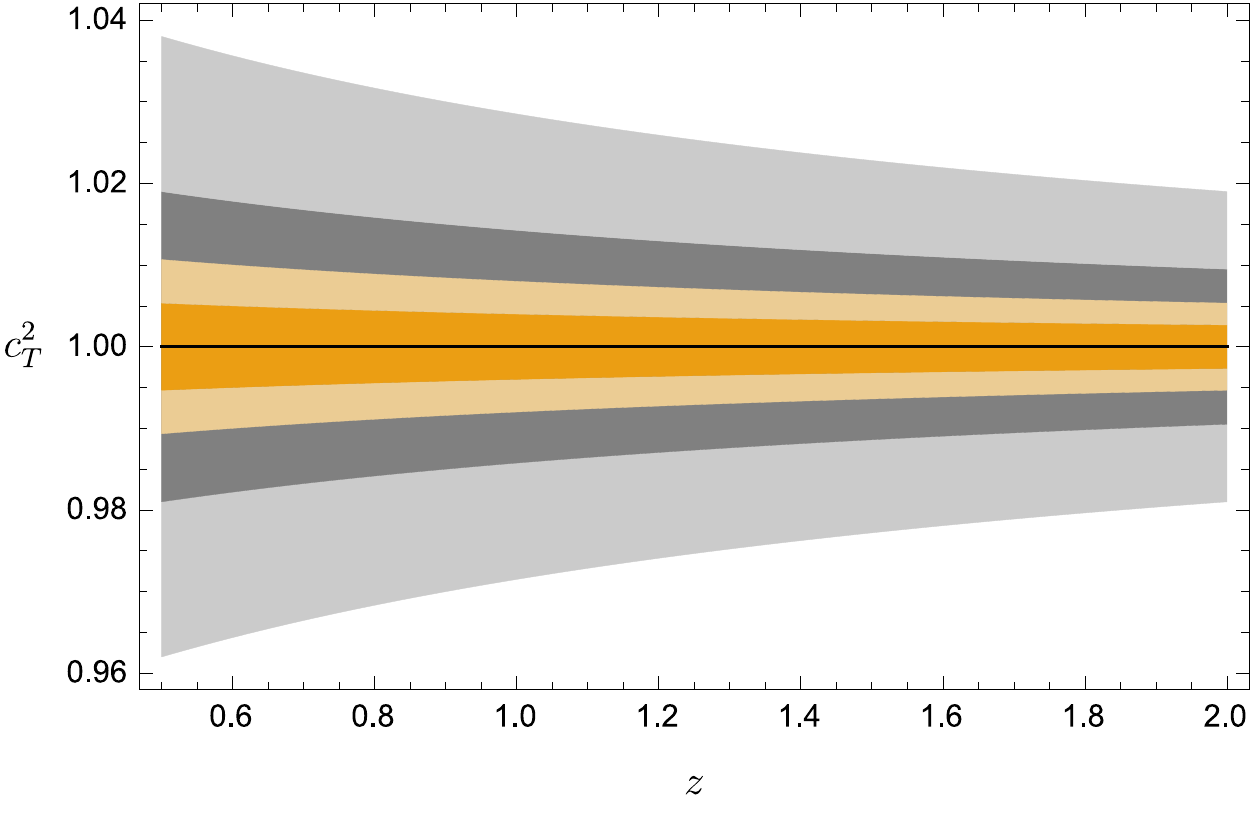}
\caption{Reconstruction of $c_T^2$ for $\alpha_{M0}<0$ (left panel) and $\alpha_{M0}>0$ (right panel) using the 68\% and 95\% C.L. results for \textit{model I} from the Fisher information analysis on the ET and DECIGO experiments. The colours match those of Figures \ref{fig:contours_alphaM0<0_Model_I} and \ref{fig:contours_alphaM0>0_Model_I}.}
\label{fig:cT_Model_I}
\end{center}
\end{figure*}

In Figure \ref{fig:cT_Model_I}, we show the reconstruction of $c_T^2$ for model I, in units of $c = 1$, as a function of the cosmic time up to $z = 2$ (statistical limit information of our mock SS catalogs). In the reconstruction, we considered $z = 0.5$ as a cut-off, in order to quantify $c_T^2$ at high $z$ only. We remind that, in our analysis, we assume $c_T/c = 1$ at low $z$, motivated by GW170917 event.  Thus, we are bounding $c_T^2$ at high $z$ in such a way to be compatible with GW170917 at very low $z$. The reconstruction is done for both experiments and, as expected, the constraints are stronger in the case of DECIGO.
The statistical bounds are insensitive to the stability condition on $\alpha_{M0}$. We find that $c_T^2$ can deviate from general relativity 3.5\% (1.8\%) at the 68\% (95\%) C.L. from ET and 1\% (0.5\%) at the 68\% (95\%) C.L. from DECIGO, with maximum deviations at intermediate $z$. These deviations tend to become smaller at larger $z$.

\begin{figure*}
\begin{center}
\includegraphics[width=3.in]{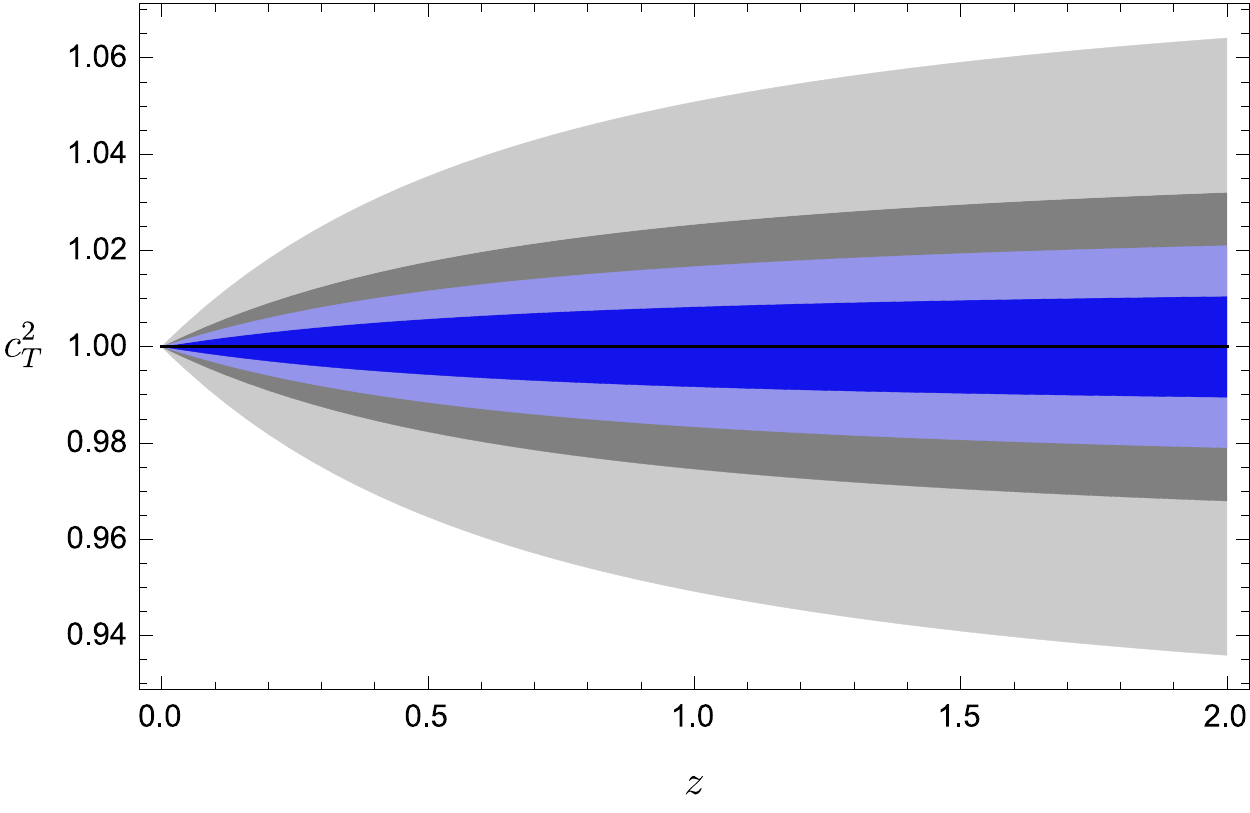} 
\includegraphics[width=3.in]{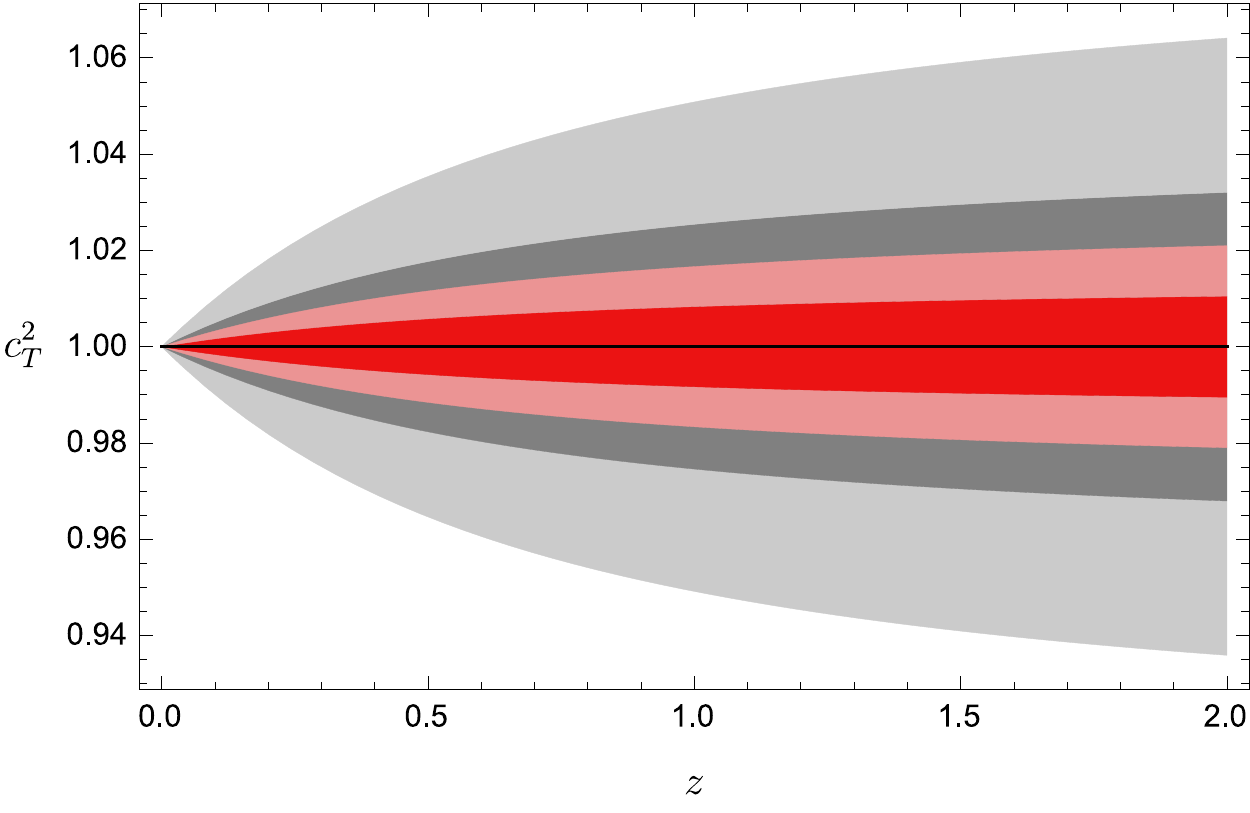}
\caption{Same as in Figure \ref{fig:cT_Model_I}, but for \textit{model II}. The colours match those of Figures \ref{fig:contours_alphaM0<0_Model_II} and \ref{fig:contours_alphaM0>0_Model_II}.}
\label{fig:cT_Model_II}
\end{center}
\end{figure*}

In Figure \ref{fig:cT_Model_II}, we show the reconstruction up to the 95\% C.L. of $c_T^2$ for model II. To analyze the main differences between the two different parametrizations, we show the reconstruction of $c_T(z)$ for model II over the whole $z$ range considered in the GWs experiments. In particular, we can see that $c_T \rightarrow 1$ for low $z$, where possible deviations from $c_T/c \neq 1$ are strongly bound and disadvantaged from both detectors sensitivities. This characteristic is not present in the parametrization commonly taken into account in the literature (model I analyzed here), where possible anomalies on the speed of GWs propagation show up at low $z$ (it can be seen intuitively in Figure \ref{fig:cT_Model_I}). Possible corrections, $c_T/c \neq 1$, in model II are manifested only at intermediate and high $z$, with a maximum deviation from general relativity of 2\% (6\%) at the 68\% (95\%) C.L. from ET and 1\% (0.5\%) at the 68\% (95\%) C.L. from DECIGO, at very high $z$. In this model, corrections of the same order of magnitude hold up to the CMB scale ($z \sim$ 1100). Thus, as a conservative estimate, we can conclude that possible anomalies on the speed of GWs under model II are about [0 - 2]\% and [0 - 0.5]\% at the 68\% C.L. from the ET and DECIGO sensitivities, respectively, from very low to very high $z$.

The dynamic feature inherent of our proposal is interesting in direct comparison with the model I. While in model II we can obtain possible deviations to be provided in the CMB scale until the strong bond imposed by the GW170917 event at very low-$z$, the same dynamics on model I is only possible with a very rigid $z$ cut-off at intermediate $z$. Thus, analyzing only its dynamic character, we can claim that our proposal is more realistic and compatible with observations.

\section{Implications on modified gravity phenomenology}
\label{Consequences}

We here briefly discuss the implications of our forecast analysis on some examples of modified gravity models. Although our bounds on  $\alpha_{T0}$ and $\alpha_{M0}$ have been obtained parametrically, without loss of generality, we can follow the opposite path and consider appropriate models in order to study consequences on some scenarios. Since model II provides a better dynamics in view of a direct comparison of the $c_T(z)$ cosmic evolution with different observations, we will take the results from this case only in our discussions in this section. However, similar conclusions can be easily drawn from the results obtained using model I. Within the Horndeski theories of gravity \cite{Horndeski, Deffayet, Kobayashi11}, the GWs propagation speed can be described by $c_T^2 = 1 + \alpha_T$ \cite{Bellini}, and one has
\begin{equation}
\label{}
M^2_{*} \alpha_T = 2 X (2 G_{4X} - 2G_{5\phi} - (\ddot{\phi} - \dot{\phi}) G_{5X})\ ,
\end{equation}
where the functions $G_{4,5}$ depend on the scalar field $\phi$ and $X \equiv -1/2 (\nabla^{\mu} \phi \nabla_{\mu} \phi)$ is the kinetic term. The function $M^2_{*}$ is the rate of evolution of the effective Planck mass. 

As a first example, we consider the derivative coupling theory in which the scalar field couples to the Einstein tensor in the form $\xi \phi G_{\mu \nu} \nabla^{\mu} \nabla^{\nu} \phi$ \cite{Amendola, Germani}. The parameter $\xi$ represents the coupling constant of the theory and quantifies possible anomalies on the GWs speed propagation. The general relativity case is recovered when $\xi = 0$. In the Horndeski Lagrangian formalism, this scenario corresponds to $G_4 = M^2_{pl}/2$ and $G_5 = \xi \phi$. Then, we can write the tensor speed excess as $M^2_{*} \alpha_T = 2 \xi \dot{\phi}^2$. Thus, the speed of GWs is given by 
\begin{equation}
c^2_T = 1 + \frac{2 \dot{\phi}^2}{M^2_{*}} \xi\ .
\end{equation}
 
 At intermediate $z$, it is reasonable to assume $\dot{\phi}/M_{*} \simeq 1$, so that we estimate $- 0.005 \lesssim \xi  \lesssim 0.005$ ($- 0.0018 \lesssim \xi  \lesssim 0.0018$) at the 95\% C.L. from ET (DECIGO), in the case $\alpha_{M0} < 0$. Similar conclusions, in terms of order of magnitude on $\xi$, are obtained when considering the stability condition $\alpha_{M0} > 0$.

A second well-known scenario where $c_T \neq c$ are the covariant Galileons models \cite{Nicolis,Deffayet02}, which are characterized by the functions $G_4 = M^2_{pl}/2 + \beta_4 X^2$ and $G_5 = \beta_5 X^2$, where $\beta_4$ and $\beta_5$ are constants. This generates a non-trivial evolution for the speed of GWs, namely 
\begin{equation}
\label{}
c^2_T = 1 + \frac{\dot{\phi}^4}{M^2_{*}} \Big[2 \beta_4 - \beta_5 (\ddot{\phi} - \dot{\phi}) \Big]\ .
\end{equation}
In the particular cases $\beta_4  = \beta_5 = 0$ and $\beta_5 = 0$, we obtain the cubic and quartic Galileon scenarios, respectively. For a qualitative discussion, we consider the quartic Galileon model, for which we find the constraint $ - 0.005 \lesssim 2 \beta_4 \dot{\phi}^4/M^2_{*} \lesssim 0.005$ at the 95\% C.L from ET sensitivity. An exact and accurate solution for $\phi(t)$ would require a thorough investigation of the cosmic dynamics, but assuming that the field does not vary much with respect to Planck mass, within the validity range of our analysis, we can estimate $ |\beta_4| \sim \mathcal{O}(10^{-3})$. In general, the full dynamics of the model should be consistent with the aforementioned range.

Many other gravity models predict possible anomalies on GWs propagation, where in general our parametric boundaries can be used to impose some limitation on such theories. It is not our main aim to make an exhaustive qualitative comparison with the most diverse scenarios. In short, our results show that for any and all models that theoretically predict $c_T/c \neq 1$, this ratio must be approximately in the range $0.97 \lesssim c_T/c \lesssim 1.03$ and $ 0.99 \lesssim c_T/c \lesssim 1.01$ at the 68\% C.L. as predicted by ET and DECIGO, respectively, at very high $z$ scale, for both stability conditions on the running of the Planck mass, when analyzed within the perspective of model II. 

\section{Final remarks}
\label{Conclusions}

Modifications of the general relativity theory are motivated mainly to explain the dark sector of the Universe. Due to extra degrees of freedom of gravitational origin, modified gravity models predict physical properties beyond the standard features of general relativity. Among several consequences, many theories call for possible anomalies on GWs propagation. The observation of the GW170817 event has imposed that the speed of GWs is equal to the speed of light for scales less than 40 Mpc. This puts strong limits and discards any model that predicts $c_T \neq c$  within this cosmological scale. However, the dark energy effects are manifested on large scales. Motivated by this aspect, we thus performed a forecast analysis using 1000 standard siren events from BNS mergers, within the sensitivity predicted for ET and DECIGO up to $z =2$ ($\sim$ 15539 Mpc). In so doing, we searched for a new bound on the ratio $c_T/c$  when considering modifications on GWs propagation between source and detector, considering corrections on the speed of GWs. We found $|c_T/c - 1| \leq 10^{-2} \,\,\, (10^{-2})$ from ET (DECIGO)
assuming model I, and $|c_T/c - 1| \leq 10^{-1} \,\,\, (10^{-2})$ from ET (DECIGO) for model II, which leaves room for small possible corrections predicted by alternative theories, compared to the only information from GW170817 event at very low $z$. Our analysis relies on a simple functional form of $\alpha_T$, and other more robust parametric functions may be used to model the $c_T(z)$ dynamics more effectively. Nevertheless, the main findings of this work represent the first forecast analysis obtained by using information from SS mock data from future detector design. In this respect, our results open a new window for possible tests on $c_T(z)$ in the future.
\\

\appendix

\section{Checking the impact of the galaxy velocity dispersion}
\label{appendix}

\begin{figure}
\begin{center}
\includegraphics[width=4.5in]{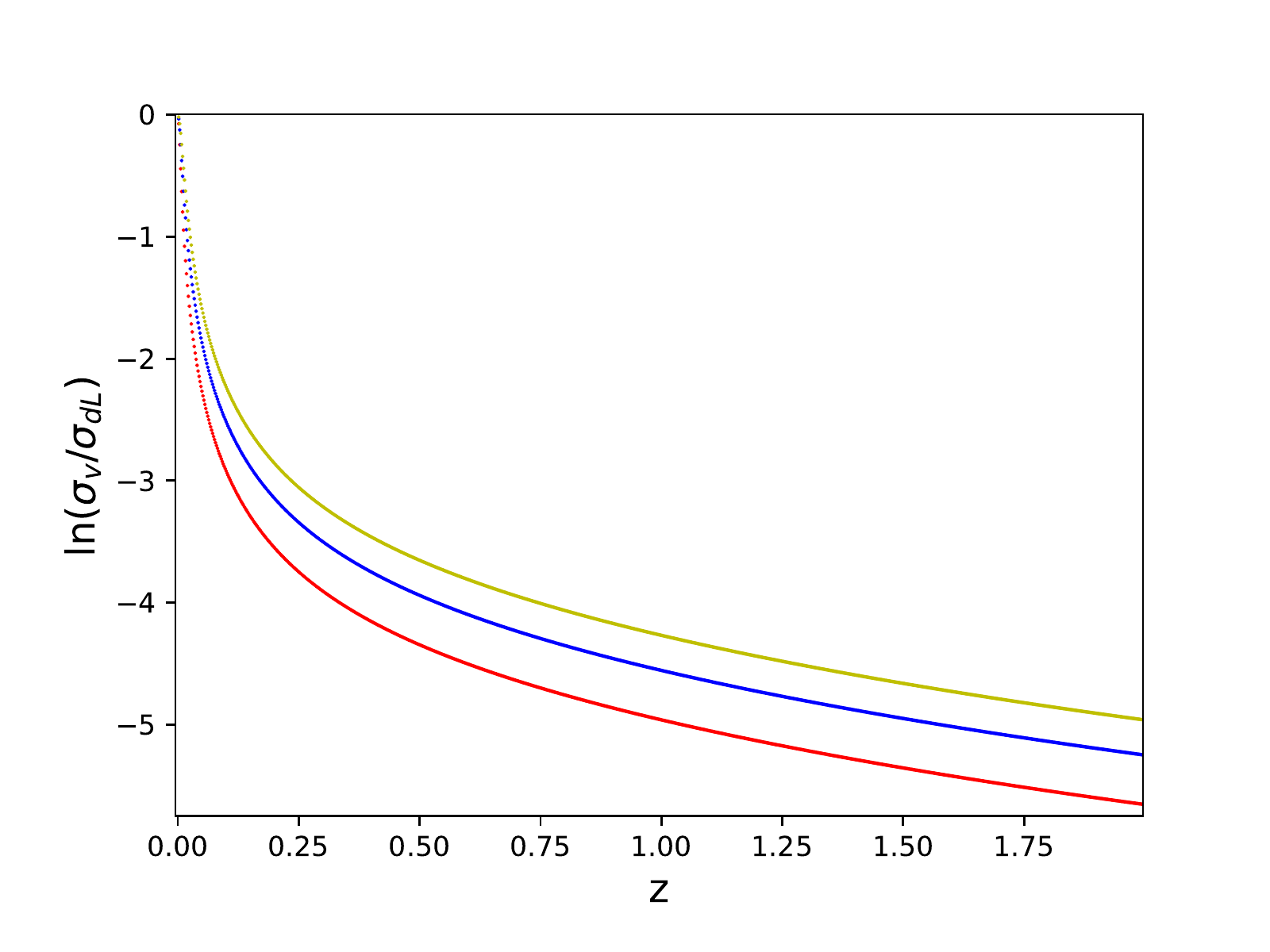}
\caption{Ratio between the galaxy peculiar velocity error ($\sigma_{\rm v, gal}$) and the total error ($\sigma_{\rm d_L}$) as a function of the redshift for different galaxy velocity dispersion values on a logarithmic scale: 200 km/s (red), 300 km/s (blue) and 400 km/s (yellow).}
\label{fig:ratio_error}
\end{center}
\end{figure}

In this Appendix, we study the impact of different galaxy velocity dispersions on our forecast analysis. As known, the galaxy peculiar velocity error has only significant influence on very-low $z$ events. Figure \ref{fig:ratio_error} quantifies this point by showing the ratio between the galaxy peculiar velocity error on the total error estimative on $d_L(z)$ as a function of $z$. We note that at high $z$, where we are particularly interested, the effects due to $\sigma_{\rm v, gal}$ should not significantly influence the forecast analysis on large cosmological distances. In fact, $\sigma_{\rm v, gal}$ only contributes on scales corresponding to $z\ll 1$, with particular significant effects at $z < 0.15$.

In Tables \ref{tab:v200}, \ref{tab:v300}, and \ref{tab:v400}, we summarize our results for model I considering different galaxy velocity dispersions: 200 km/s, 300 km/s and 400 km/s. As expected, we did not notice significant changes in our main results when considering different $\sigma_{\rm v, gal}$ within the range [200 - 400] km/s, being all bounds compatible with each other. Furthermore, since we considered the $\Lambda$CDM model previsions for $z < 0.10$ in order to be locally compatible with the GW170817 event at very low $z$, in the region where we have any possible effects from $\sigma_\text{v, gal}$,\emph{ i.e.} for $z \ll 1$, we have the same dynamics as predicted by general relativity. Without loss of generality, the same arguments can be applied when considering the model II.
\\

\begin{table}[h!]
\small
\begin{center}
\begin{tabular}{c| c  c| c c}
\hline
\hline
& & Stability condition $\alpha_{M0}<0$ & Stability condition $\alpha_{M0} > 0$ \\
\hline
Parameter & $\sigma$(ET) & $\sigma$(DECIGO) & $\sigma$(ET) & $\sigma$(DECIGO) \\
\hline
\hline
$\alpha_{T0}$  & 0.054 &  0.016  & 0.056 &  0.018 \\
$\alpha_{M0}$  & $> -0.18$ &   $> -0.054$ & $< 0.089 $ & $< 0.030$  \\
$n_{1}$        & 0.40 &  0.13   & 0.33 &  0.11 \\
$n_{2}$        & 0.23 &  0.061  & 0.22 & 0.060  \\
\hline
\hline
\end{tabular}
\caption{Forecast constraints from the ET and DECIGO experiments, assuming velocity dispersion of the galaxy set to be 200 km/s and under both stability condition $\alpha_{M0}<0, >0$ for all free parameter of the theory. The notation is the same as in Table \ref{tab:results_Model_I}.}
\label{tab:v200}
\end{center}
\end{table}


\begin{table}[h!]
\small
\begin{center}
\begin{tabular}{c| c  c| c c }
\hline
\hline
& & Stability condition $\alpha_{M0}<0$ & Stability condition $\alpha_{M0} > 0$ \\
\hline
Parameter & $\sigma$(ET) & $\sigma$(DECIGO) & $\sigma$(ET) & $\sigma$(DECIGO) \\
\hline
$\alpha_{T0}$  & 0.053 &  0.015 &  0.057 &  0.016  \\
$\alpha_{M0}$  & $>$ -0.18 &   $>$ -0.050  & $<$ 0.087 & $<$ 0.052   \\
$n_{1}$        & 0.41 &  0.13  & 0.33 &  0.12 \\
$n_{2}$        & 0.18 &  0.053  & 0.17 &  0.052 \\
\hline
\hline
\end{tabular}
\caption{Same as in Table \ref{tab:v200}, but assuming velocity dispersion of the galaxy set to be 300 km/s.}
\label{tab:v300}
\end{center}
\end{table}

\begin{table}[h!]
\small
\begin{center}
\begin{tabular}{c| c  c| c c }
\hline
\hline
& & Stability condition $\alpha_{M0}<0$ & Stability condition $\alpha_{M0} > 0$ \\
\hline
Parameter & $\sigma$(ET) & $\sigma$(DECIGO) & $\sigma$(ET) & $\sigma$(DECIGO) \\
\hline
$\alpha_{T0}$  & 0.055 &  0.014 & 0.060 &  0.016  \\
$\alpha_{M0}$  & $> -0.19$ &   $> -0.050$  & $< 0.090 $ & $< 0.024$   \\
$n_{1}$        & 0.45 &  0.13 & 0.34 &  0.11   \\
$n_{2}$        & 0.22 &  0.051 & 0.21 & 0.049   \\
\hline
\hline
\end{tabular}
\caption{Same as in Table \ref{tab:v200}, but assuming velocity dispersion of the galaxy set to be 400 km/s.}
\label{tab:v400}
\end{center}
\end{table}

\begin{acknowledgments}
\noindent  The authors thank the referee for his/her valuable comments and suggestions. RD acknowledges the support of INFN (iniziativa specifica QGSKY). RCN would like to thank the Brazilian agency FAPESP for financial support under Project No. 2018/18036-5. JCNA would like to thank FAPESP for financial support under Project No. 2013/26258-4, and CNPq for partial financial support under Grant No. 307217/2016-7.

\end{acknowledgments}

\end{document}